\documentclass[journal]{IEEEtran}

\usepackage{graphicx}
\usepackage{epstopdf}
\usepackage{verbatim}
\usepackage{times}
\usepackage{amsfonts}
\usepackage{subfigure}
\usepackage{url}
\usepackage[]{amsmath}
\usepackage[numbers,sort&compress]{natbib}
\usepackage{stfloats}
\usepackage{bm}

\begin{document}
\newcounter{mytempeqncnt}
\title{Antenna Array Based Positional Modulation with a Two-Ray Multi-Path Model}
\author{\IEEEauthorblockN{Bo Zhang and Wei Liu}\\
\IEEEauthorblockA{Communications Research Group, Department of Electronic and Electrical Engineering\\University of Sheffield, Sheffield S1 4ET, United Kingdom}}
\maketitle
\begin{abstract}
Traditional directional modulation (DM) designs are based on the assumption that there is no multi-path effect between transmitters and receivers. One problem with these designs is that the resultant systems will be vulnerable to eavesdroppers which are aligned with or very close to the desired directions, as the received modulation pattern at these positions is similar to the given one. To solve the problem, a two-ray multi-path model is studied for positional modulation and the coefficients design problem for a given array geometry and a location-optimised antenna array is solved, where the multi-path effect is exploited to generate a given modulation pattern at desired positions, with scrambled values at positions around them.
 \end{abstract}



%

\section{Introduction}
Directional modulation (DM), as a security technique to keep known constellation mappings in a desired direction or directions, while scrambling them for the remaining ones, was introduced in~\cite{babakhani09a} by combining the direct radiation beam and reflected beams in the far-field. In~\cite{daly10a}, a reconfigurable array was designed by switching elements for each symbol to make their constellation points not scrambled in desired directions, but distorted in other directions. A method named dual beam DM was introduced in~\cite{hong11a}, where the I and Q signals are transmitted by different antennas. In~\cite{daly09a,zhang17a}, phased arrays were employed to show that DM can be implemented by phase shifting the transmitted antenna signals properly. Multi-carrier based phased antenna array design for directional modulation was studied in~\cite{zhang18b}, followed by a combination of DM and polarisation design in~\cite{zhang18a}. The bit error rate (BER) performance of a system based on a two-antenna array was studied using the DM technique for eight phase shift keying modulation in~\cite{shi13a}. A more systematic pattern synthesis approach was presented in~\cite{ding13b}, followed by a time modulation technique for DM to form a four-dimensional (4-D) antenna array in~\cite{zhu14a}.

However, eavesdroppers aligned with or very close to the desired direction/directions will be a problem for secure signal transmission, as their received modulation patterns are similar to the given one. To make sure that a given modulation pattern can only be received at certain desired positions, one solution is adopting a multi-path model, where signals via both line of sight (LOS) and reflected paths are combined at the receiver side~\cite{shi13b,ding15a,ding16a,kalantari16a,ding17a}. In this work, the typical two-ray multi-path model is further studied based on an antenna array and a closed-form solution is provided. Such a two-ray model is more realistic in the millimetre wave band given the more directional propagation model in this frequency band. Furthermore, the antenna location optimisation problem is investigated in the context of positional modulation and a compressive-sensing based design is proposed.

The remaining part of this paper is structured as follows. A review of the two-ray model is given in Sec. \ref{sec:Review of Two-Path Model}. Positional modulation design based on a given array geometry and an array with optimised antenna locations are presented in Sec. \ref{sec:Proposed design for Positional modulation design}. Design examples are provided in Sec. \ref{sec:sim}, followed by conclusions in Sec. \ref{sec:con}.

\section{Review of Two-Path Model}\label{sec:Review of Two-Path Model}

\begin{figure}
  \centering
 \includegraphics[width=0.45\textwidth]{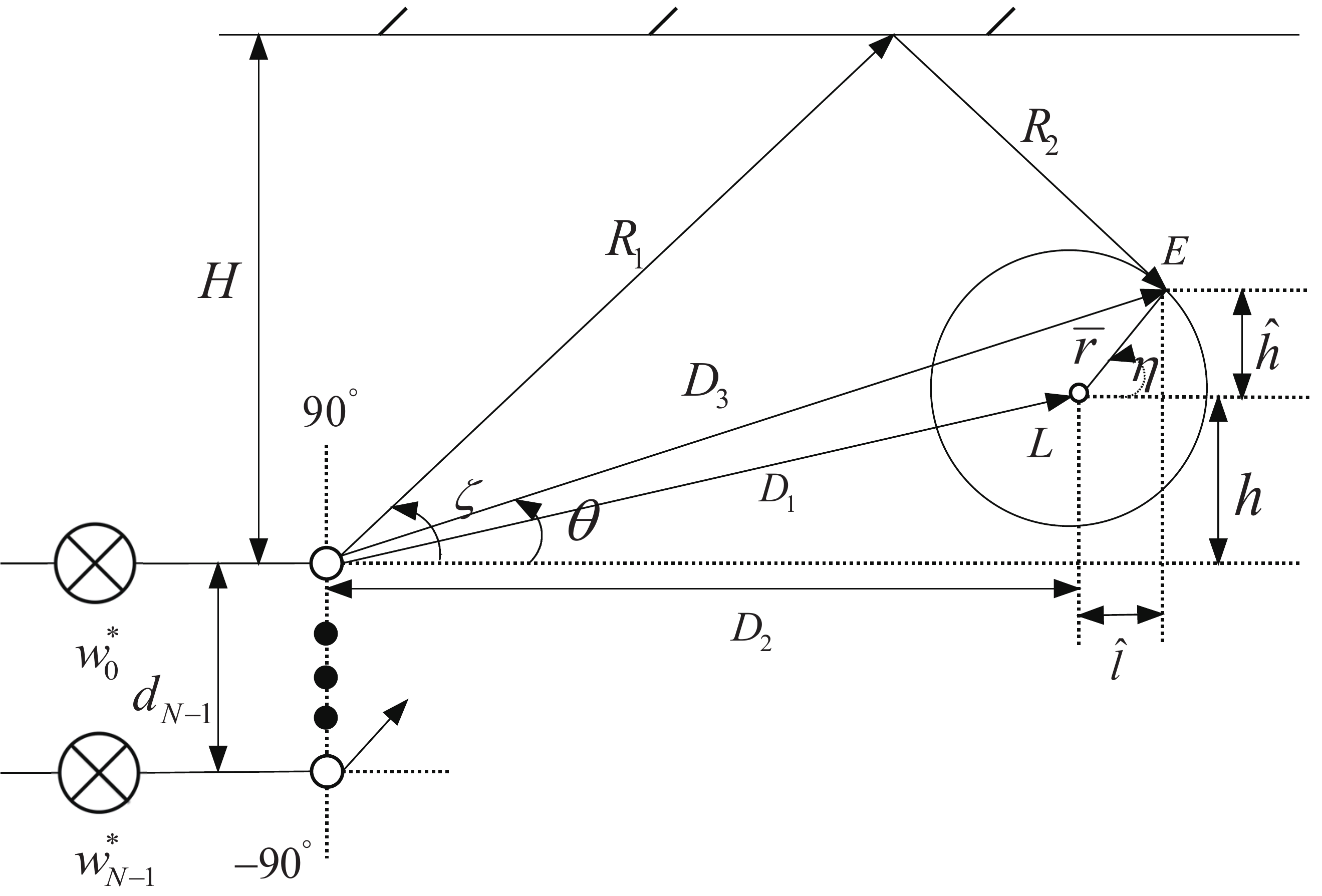}\\
  \caption{Multi-path signal transmission to the desired receiver $L$ and eavesdroppers $E$}\label{fig:narrowband_multipath_legit_and_eavesdroppers_360}
\end{figure}
An $N$-element omni-directional linear antenna array for transmit beamforming~\cite{shi13b} is shown in Fig. \ref{fig:narrowband_multipath_legit_and_eavesdroppers_360}, where the spacing between the zeroth and the $n$-th antennas is represented by $d_n$ for $n = 1,\ldots, N-1$, with the transmission angle $\theta \in [-90^{\circ},90^{\circ}]$. The weight coefficient of each antenna is denoted by $w_{n}$, for $n=0, \ldots, N-1$. The desired position is represented by $L$ with a distance $D_1$ to the transmit array and a vertical distance $h$ to the broadside direction where $h$ is positive for $L$ above the broadside direction and negative for the opposite. The projection of $D_1$ onto the broadside direction is represented by $D_2$. The positions of eavesdroppers $E$ are shown on the circumference of the circle, with the radius $\bar{r}$ and angle $\eta\in[0^\circ,360^\circ)$ to the circle centre $L$. For eavesdroppers in the direction $\eta$, we have the corresponding $\hat{h}$ and $\hat{l}$, representing the vertical height and horizontal length relative to the centre point L, with $\bar{r} = \sqrt{\hat{h}^2+\hat{l}^2}$, and the distance to transmitters is represented by $D_3$. To produce the required reflected path, a reflecting surface with a distance $H$ above and perpendicular to the antenna array is created to form the two-ray model. The reflected distances $R_1$ and $R_2$ represent the path length before and after reflection, and the transmission angle for the reflected path is determined by $\zeta \in (0^{\circ},90^{\circ}]$.

For signals transmitted to the desired location $L$, as shown in Fig. \ref{fig:narrowband_multipath_legit_and_eavesdroppers_360}, we have
\begin{equation}
\begin{split}
D_2 &= \sqrt{D_1^2-h^2},
\quad \theta = \tan^{-1}(h/D_2),\\
\zeta &= \tan^{-1}((2H-h)/D_2).
\end{split}
\end{equation}
For signals transmitted to the eavesdroppers,
\begin{equation}
\label{eq:relationships between d1 and d3}
\begin{split}
   \hat{h}(\eta) &= \bar{r}\sin\eta,\quad
   \hat{l}(\eta) = \bar{r}\cos\eta,\\
   D_3 &= \sqrt{(D_2+\hat{l})^2+(h+\hat{h})^2}.
\end{split}
\end{equation}
The corresponding $\theta(\eta)$ and $\zeta(\eta)$ for the LOS and reflected paths can be formulated as
\begin{equation}
\label{eq:theta value for eavesdroppers}
\begin{split}
\theta(\eta) &= \tan^{-1}((h+\hat{h})/(D_2+\hat{l})),\\
\zeta(\eta) &= \tan^{-1}((2H-\hat{h}-h)/(D_2+\hat{l})).
\end{split}
\end{equation}
Then, for the reflected path, $R_1(\zeta)$ and $R_2(\zeta)$ are given by
\begin{equation}
R_1(\zeta) = H/\sin\zeta,\quad
R_2(\zeta) = (H-h-\hat{h})/\sin\zeta.
\end{equation}
The steering vector for the LOS path and the reflected path in two-ray model are, respectively, given by
\begin{equation}
\begin{split}
    \textbf{s}(\omega,\theta) &= [1, e^{j\omega d_1\sin\theta/c}, \ldots, e^{j\omega d_{N-1}\sin\theta/c}]^{T},\\
    \hat{\textbf{s}}(\omega,\zeta) &= [1, e^{j\omega d_1\sin\zeta/c}, \ldots, e^{j\omega d_{N-1}\sin\zeta/c}]^{T}.
\end{split}
\end{equation}

Moreover, phase shift and power attenuation caused by these multiple paths need to be considered~\cite{shi13b}. When $\hat{h}$ and $\hat{l}$ are both zero-valued, as shown in \eqref{eq:relationships between d1 and d3}, $D_3 = D_1$. Therefore, we can consider the length $D_1$ as a special case of the length $D_3$. Then the phase shifts for LOS paths is given by
\begin{equation}
\psi(\theta) = 2\pi\times rem(D_3(\theta),\lambda),
\end{equation}
where $rem(A,\lambda)$ represents the remainder of $A$ divided by $\lambda$. The phase shift for the reflected path is determined by $R_1(\zeta)+R_2(\zeta)$ and given by
\begin{equation}
\phi(\zeta) = \pi+2\pi\times rem(R_1(\zeta)+R_2(\zeta),\lambda),
\end{equation}
where $\pi$ is caused by the reflecting surface. The attenuation ratio for a LOS is given by~\cite{shi13b}
\begin{equation}
\nu(\theta) = D/D_3(\theta).
\end{equation}
Here $D$ is assumed to be the distance where the received signal has unity power. Similarly, the attenuation ratio for the signal received via the reflected path is given by
\begin{equation}
\xi(\zeta) = D/(R_1(\zeta)+R_2(\zeta)).
\end{equation}

Then, in the two-ray model, the beam response of the array, represented by $p(\theta,\zeta)$, is a combination of signals through the LOS path and the reflected path,
\begin{equation}
\begin{split}
    &p(\theta,\zeta)=\\
    &\nu(\theta)e^{j\psi(\theta)}(\textbf{w}^{H}\textbf{s}(\omega,\theta))
    +\xi(\zeta)e^{j\phi(\zeta)}(\textbf{w}^{H}\hat{\textbf{s}}(\omega,\zeta)),
\end{split}
\end{equation}
with the weight vector $\textbf{w} = [w_{0}, w_{1}, \ldots, w_{N-1}]^{T}$.

\section{Proposed design for Positional modulation design}\label{sec:Proposed design for Positional modulation design}
\subsection{Positional modulation design for a given array geometry}

The objective of positional modulation design is to find a set of weight coefficients creating signals with a given modulation pattern to desired locations, while the modulations of the signals received around them are distorted. For $M$-ary signaling, such as multiple phase shift keying (MPSK), there are $M$ sets of desired array responses $p_m(\theta,\zeta)$, with a corresponding weight vector $\textbf{w}_{m}=[w_{0,m}, \ldots, w_{N-1,m}]^T$, $m=0, \ldots, M-1$. Assuming in total $R$ locations in the design ($r$ desired locations and $R-r$ eavesdropper locations), we can have the corresponding transmission angles $\theta_k$ for LOS and $\zeta_k$ for the reflected path to the $k$-th position, $k = 0,\ldots,R-1$. Then an $N\times r$ matrix $\textbf{S}_{L}$ is constructed as the set of steering vectors for the LOS path to desired receivers, and similarly we have $\textbf{S}_{E} = [\textbf{s}(\omega,\theta_0),\textbf{s}(\omega,\theta_1), \ldots, \textbf{s}(\omega,\theta_{R-r-1})]$ (an $N\times (R-r)$ matrix) for steering vectors to eavesdroppers. The corresponding steering vectors for the reflected path to desired receivers and eavesdroppers are given by $\hat{\textbf{S}}_{L}$ and $\hat{\textbf{S}}_{E}$, respectively. $\textbf{p}_{m,L}$ ($1\times r$ vector) and $\textbf{p}_{m,E}$ ($1\times (R-r)$ vector) are required responses for the desired locations and the eavesdroppers for the $m$-th constellation point.

Moreover, the phase shifts for the LOS and reflected paths to both eavesdroppers and desired receivers, and their corresponding attenuation ratios are given by
\begin{equation}
\begin{split}
\bm{\psi}_{E} & = [\psi(\theta_0), \psi(\theta_1), \ldots, \psi(\theta_{R-r-1})],\\
\bm{\psi}_{L} & = [\psi(\theta_{R-r}), \psi(\theta_{R-r+1}), \ldots, \psi(\theta_{R-1})],\\
\bm{\phi}_{E} & = [\phi(\zeta_0), \phi(\zeta_1), \ldots, \phi(\zeta_{R-r-1})],\\
\bm{\phi}_{L} & = [\phi(\zeta_{R-r}), \phi(\zeta_{R-r+1}), \ldots, \phi(\zeta_{R-1})],\\
\bm{\nu}_{E} & = [\nu(\theta_0), \nu(\theta_1), \ldots, \nu(\theta_{R-r-1})],\\
\bm{\nu}_{L} & = [\nu(\theta_{R-r}), \nu(\theta_{R-r+1}), \ldots, \nu(\theta_{R-1})],\\
\bm{\xi}_{E} & = [\xi(\zeta_0), \xi(\zeta_1), \ldots, \xi(\zeta_{R-r-1})],\\
\bm{\xi}_{L} & = [\xi(\zeta_{R-r}), \xi(\zeta_{R-r+1}), \ldots, \xi(\zeta_{R-1})].
 \end{split}
\end{equation}

Then, for the $m$-th constellation point, the coefficients can be formulated as
\begin{equation}
\begin{split}
\label{eq:mindm}
    &\underset{\textbf{w}_m}{\text{min}}
    ||\textbf{p}_{m,E}-(\bm{\nu}_{E}\cdot e^{j\bm{\psi}_{E}}\cdot(\textbf{w}_{m}^{H}\textbf{S}_{E})+\bm{\xi}_{E}\cdot e^{j\bm{\phi}_{E}}\cdot(\textbf{w}_m^{H}\hat{\textbf{S}}_{E}))||_{2}\\
    &\text{subject to}\\
    &\bm{\nu}_{L}\cdot e^{j\bm{\psi}_{L}}\cdot (\textbf{w}_{m}^{H}\textbf{S}_{L})+\bm{\xi}_{L}\cdot e^{j\bm{\phi}_{L}}\cdot(\textbf{w}_m^{H}\hat{\textbf{S}}_{L}) = \textbf{p}_{m,L},
\end{split}
\end{equation}
where $\cdot$ is the dot product. Its solution can be solved by the method of Lagrange multipliers, and the optimum value for the weight vector $\textbf{w}_{m}$ is given by
\begin{equation}
\begin{split}
\textbf{w}_{m} = &\textbf{K}_5^{-1}(\hat{\textbf{S}}_{E}\textbf{K}_2\textbf{p}^H_{m,E}-\textbf{S}_{E}\textbf{K}_1\textbf{p}^H_{m,E}\\
&-\textbf{K}_6^H\textbf{S}_{L}\textbf{K}_3-\textbf{K}_6^H\hat{\textbf{S}}_{L}\textbf{K}_4)
\end{split}
\end{equation}
where
\begin{equation}
\begin{split}
\textbf{K}_1 &= diag(\bm{\nu}_{E}diag(e^{j\bm{\psi}_{E}})),\quad \textbf{K}_2 = diag(\bm{\xi}_{E}diag(e^{j\bm{\phi}_{E}})),\\
\textbf{K}_3 &= diag(\bm{\nu}_{L}diag(e^{j\bm{\psi}_{L}})),\quad \textbf{K}_4 = diag(\bm{\xi}_{L}diag(e^{j\bm{\phi}_{L}})),\\
\textbf{K}_5 &= \textbf{S}_{E}\textbf{K}_1\textbf{K}_1^H\textbf{S}^H_{E}+\textbf{S}_{E}\textbf{K}_1\textbf{K}_2^H\textbf{S}^H_{E}\\
&+\hat{\textbf{S}}_{E}\textbf{K}_2\textbf{K}^H_1\textbf{S}^H_{E}+\hat{\textbf{S}}_{E}\textbf{K}_2\textbf{K}^H_2\hat{\textbf{S}}^H_{E},\\
\textbf{K}_6 &= (\textbf{p}_{m,E}\textbf{K}^H_2\hat{\textbf{S}}^H_{E}\textbf{K}_5^{-H}\textbf{S}_{L}\textbf{K}_3-\textbf{p}_{m,E}\textbf{K}^H_1\textbf{S}^H_{E}\textbf{K}_5^{-H}\textbf{S}_{L}\textbf{K}_3\\
&-\textbf{p}_{m,E}\textbf{K}^H_2\hat{\textbf{S}}^H_{E}\textbf{K}_5^{-H}\hat{\textbf{S}}_{L}\textbf{K}_4-\textbf{p}_{m,E}\textbf{K}^H_1\textbf{S}^H_{E}\textbf{K}_5^{-H}\hat{\textbf{S}}_{L}\textbf{K}_4\\
&-\textbf{p}_{m,L})\\
&\times(\textbf{K}_3^H\textbf{S}^H_{L}\textbf{K}_5^{-H}\textbf{S}_{L}\textbf{K}_3+\textbf{K}_4^H\hat{\textbf{S}}^H_{L}\textbf{K}_5^{-H}\textbf{S}_{L}\textbf{K}_3\\
&+\textbf{K}_3^H\textbf{S}^H_{L}\textbf{K}_5^{-H}\hat{\textbf{S}}_{L}\textbf{K}_4+\textbf{K}_4^H\hat{\textbf{S}}^H_{L}\textbf{K}_5^{-H}\hat{\textbf{S}}_{L}\textbf{K}_4)^{-1}.
\end{split}
\end{equation}

\subsection{Positional modulation design for an optimised locations array}

Equation \eqref{eq:mindm} is for designing the positional modulation coefficients for a given set of antenna locations. In practice, we may opt to find optimised locations to construct an array for an improved performance, which can be considered as a sparse antenna array design problem~\cite{moffet68a,vantrees02a}. Many methods have been proposed for the design of a general sparse antenna array, including the genetic algorithm~\cite{haupt94a,yan97a,cen12a},
simulated annealing~\cite{trucco99a}, and compressive sensing (CS)~\cite{prisco11a,carin09a,oliveri12a,hawes14a}, and in this section, CS-based methods is studied.

For CS-based sparse array design for positional modulation, a given aperture is densely sampled with a large number ($N$) of potential antennas, as shown in Fig \ref{fig:narrowband_multipath_legit_and_eavesdroppers_360}, and the values of $d_{n}$, for $n=1, 2, \ldots, N-1$, are selected to give a uniform grid. Through selecting the minimum number of non-zero valued weight coefficients, where the corresponding antennas are kept, and the rest of the antennas with zero-valued coefficients are removed, to generate a response close to the desired one, sparseness of the design is acquired~\cite{zhang17a,zhang18b}. Then for the $m$-th constellation point, the cost function is $\underset{\textbf{w}_m}{\text{min}}||\textbf{w}_{m}||_1$ and the constraints are $||\textbf{p}_{m,E}-(\bm{\nu}_{E}\cdot e^{j\bm{\psi}_{E}}\cdot(\textbf{w}_{m}^{H}\textbf{S}_{E})+\bm{\xi}_{E}\cdot e^{j\bm{\phi}_{E}}\cdot(\textbf{w}_m^{H}\hat{\textbf{S}}_{E}))||_{2}\leq \alpha$ and $\bm{\nu}_{L}\cdot e^{j\bm{\psi}_{L}}\cdot (\textbf{w}_{m}^{H}\textbf{S}_{L})+\bm{\xi}_{L}\cdot e^{j\bm{\phi}_{L}}\cdot(\textbf{w}_m^{H}\hat{\textbf{S}}_{L}) = \textbf{p}_{m,L}$, where $||\cdot||_1$ is the $l_1$ norm, used as an approximation to the $l_0$ norm and $\alpha$ is the allowed difference between the desired and designed responses. As each antenna element corresponds to $M$ weight coefficients and these $M$ coefficients correspond to $M$ symbols, to remove the $n$-th antenna, we need all coefficients in the following vector $\tilde{\textbf{w}}_{n}$ to be zero-valued or $||\tilde{\textbf{w}}_{n}||_2 = 0$~\cite{zhang17a,zhang18b},
\begin{equation}
\tilde{\textbf{w}}_{n}=[w_{n,0}, \ldots, w_{n,M-1}],
\end{equation}
where $w_{n,m}$ represents the coefficients on the $n$-th antenna for the $m$-th symbol. Then, to calculate the minimum number of antenna elements, we gather all $||\tilde{\textbf{w}}_{n}||_2$ for $n = 0,\ldots, N-1$ to form a new vector $\hat{\textbf{w}}$,
\begin{equation}
\hat{\textbf{w}} = [||\tilde{\textbf{w}}_{0}||_{2}, ||\tilde{\textbf{w}}_{1}||_2, \ldots, ||\tilde{\textbf{w}}_{N-1}||_2]^{T}.
\end{equation}
Moreover, we need to impose positional modulation constraints including
\begin{equation}
\begin{split}
\textbf{W} &= [\textbf{w}_{0}, \textbf{w}_{1}, \ldots, \textbf{w}_{M-1}],
\textbf{P}_{E} = [\textbf{p}_{0,E}, \textbf{p}_{1,E}, \ldots, \textbf{p}_{M-1,E}]^T,\\
\textbf{P}_{L} &= [\textbf{p}_{0,L}, \textbf{p}_{1,L}, \ldots, \textbf{p}_{M-1,L}]^T,\\
\tilde{\bm{\nu}}_{E} &= \bm{\nu}_{E}\otimes \text{ones}(M,1),\;
\tilde{\bm{\nu}}_{L} = \bm{\nu}_{L}\otimes \text{ones}(M,1),\\
\tilde{\bm{\xi}}_{E} &= \bm{\xi}_{E}\otimes \text{ones}(M,1),\;
\tilde{\bm{\xi}}_{L} = \bm{\xi}_{L}\otimes \text{ones}(M,1),\\
\tilde{\bm{\psi}}_{E} &= \bm{\psi}_{E}\otimes \text{ones}(M,1),\;
\tilde{\bm{\psi}}_{L} = \bm{\psi}_{L}\otimes \text{ones}(M,1),\\
\tilde{\bm{\phi}}_{E} &= \bm{\phi}_{E}\otimes \text{ones}(M,1),\;
\tilde{\bm{\phi}}_{L} = \bm{\phi}_{L}\otimes \text{ones}(M,1),
\end{split}
\end{equation}
where $\otimes$ stands for the Kronecker product, and $\text{ones}(M,1)$ is an $M\times 1$ matrix of ones. Then the group sparsity based sparse array design for DM~\cite{zhang17a,zhang18b} can be formulated as
\begin{equation}
\begin{split}
\label{eq:grsp}
    &\underset{\textbf{W}}{\text{min}}||\hat{\textbf{w}}||_{1}\\
    &\text{subject to}\\
    &||\textbf{P}_{E}-(\tilde{\bm{\nu}}_{E}\cdot e^{j\tilde{\bm{\psi}}_{E}}\cdot(\textbf{W}^{H}\textbf{S}_{E})+\tilde{\bm{\xi}}_{E}\cdot e^{j\tilde{\bm{\phi}}_{E}}\cdot(\textbf{W}^{H}\hat{\textbf{S}}_{E}))||_{2}\leq \alpha\\
    &\tilde{\bm{\nu}}_{L}\cdot e^{j\tilde{\bm{\psi}}_{L}}\cdot(\textbf{W}^{H}\textbf{S}_{L})+\tilde{\bm{\xi}}_{L}\cdot e^{j\tilde{\bm{\phi}}_{L}}\cdot(\textbf{W}^{H}\hat{\textbf{S}}_{L}) = \textbf{P}_{L}.
\end{split}
\end{equation}

As the reweighted $l_1$ norm minimisation has a closer approximation to the $l_0$ norm~\cite{candes08a,prisco12a,fuchs12a}, we can further modify \eqref{eq:grsp} into the reweighted form in a similar way as in~\cite{zhang17a}, where at the $u$-th iteration,
\begin{equation}
\begin{split}
\label{eq:approx}
    &\underset{\textbf{W}}{\text{min}}\sum\limits_{n=0}^{N-1} \delta_{n}^u||\tilde{\textbf{w}}_{n}^u||_2\\
    &\text{subject to}\quad ||\textbf{P}_{E}-(\tilde{\bm{\nu}}_{E}\cdot e^{j\tilde{\bm{\psi}}_{E}}\cdot((\textbf{W}^{u})^H\textbf{S}_{E})\\
    &+\tilde{\bm{\xi}}_{E}\cdot e^{j\tilde{\bm{\phi}}_{E}}\cdot((\textbf{W}^{u})^H\hat{\textbf{S}}_{E}))||_{2}\leq \alpha\\
    &\tilde{\bm{\nu}}_{L}\cdot e^{j\tilde{\bm{\psi}}_{L}}\cdot((\textbf{W}^{u})^H\textbf{S}_{L})+\tilde{\bm{\xi}}_{L}\cdot e^{j\tilde{\bm{\phi}}_{L}}\cdot((\textbf{W}^{u})^H\hat{\textbf{S}}_{L}) = \textbf{P}_{L}.
\end{split}
\end{equation}
Here the superscript $u$ indicates the $u$-th iteration, and $\delta_{n}$ is the reweighting term for the $n$-th row of coefficients, given by $\delta_{n}^u=(||\tilde{\textbf{w}}_{n}^{u-1}||_2+\gamma)^{-1}$. ($\gamma>0$ is required to provide numerical stability and the iteration process is described as in~\cite{zhang17a}.) The problem in \eqref{eq:grsp} and \eqref{eq:approx} can be solved by cvx~\cite{cvx,grant08a}.

\section{Design examples}\label{sec:sim}

In this section, we provide several representative design examples to show the performance of the proposed formulations in the two-ray model. Without loss of generality, we assume there is one desired location at the circle centre with $\theta = 0^{\circ}$, and $H = 500\lambda$, $D_1 = D = 1000\lambda$. Eavesdroppers are located at the circumference of the circle with $\bar{r} = 8.4\lambda$ and $\eta \in [0^{\circ}, 360^{\circ})$, sampled every $1^{\circ}$. With the radius $\bar{r}$ and the angle $\eta$ based on \eqref{eq:theta value for eavesdroppers}, it can be seen that all eavesdroppers are in the directions of $\theta\in(-0.5^\circ, 0.5^\circ)$, i.e. aligned with or very close to the desired user. The desired response is a value of one magnitude (the gain is $0$dB) with $90^\circ$ phase shift at the desired location (QPSK), i.e. symbols `00', `01', `11', `10' correspond to $45^{\circ}$, $135^{\circ}$, $-135^{\circ}$ and $-45^{\circ}$, respectively, and a value of $0.1$ (magnitude) with random phase shifts at eavesdroppers. Moreover the bit error rate (BER) result is also presented. Here the signal to noise ratio (SNR) is set at $12$ dB at the desired location, and we assume the additive white Gaussian noise (AWGN) level is at the same level for all eavesdroppers.

The number of antenna elements for the ULA design is $N = 30$, while for the sparse array design, the maximum aperture of the array is set to $20\lambda$ with $401$ equally spaced potential antennas. To make a fair comparison, we use the value of error norm between desired and designed array responses calculated from the ULA design \eqref{eq:mindm} as the threshold $\alpha$ for the sparse array design. $\gamma = 0.001$ used in the reweighted $l_1$ norm minimisation \eqref{eq:approx} indicates that antennas associated with a weight value smaller than $0.001$ will be removed.

\begin{figure}
  \centering
  \subfigure[]{
  \includegraphics[width = 0.35\textwidth]{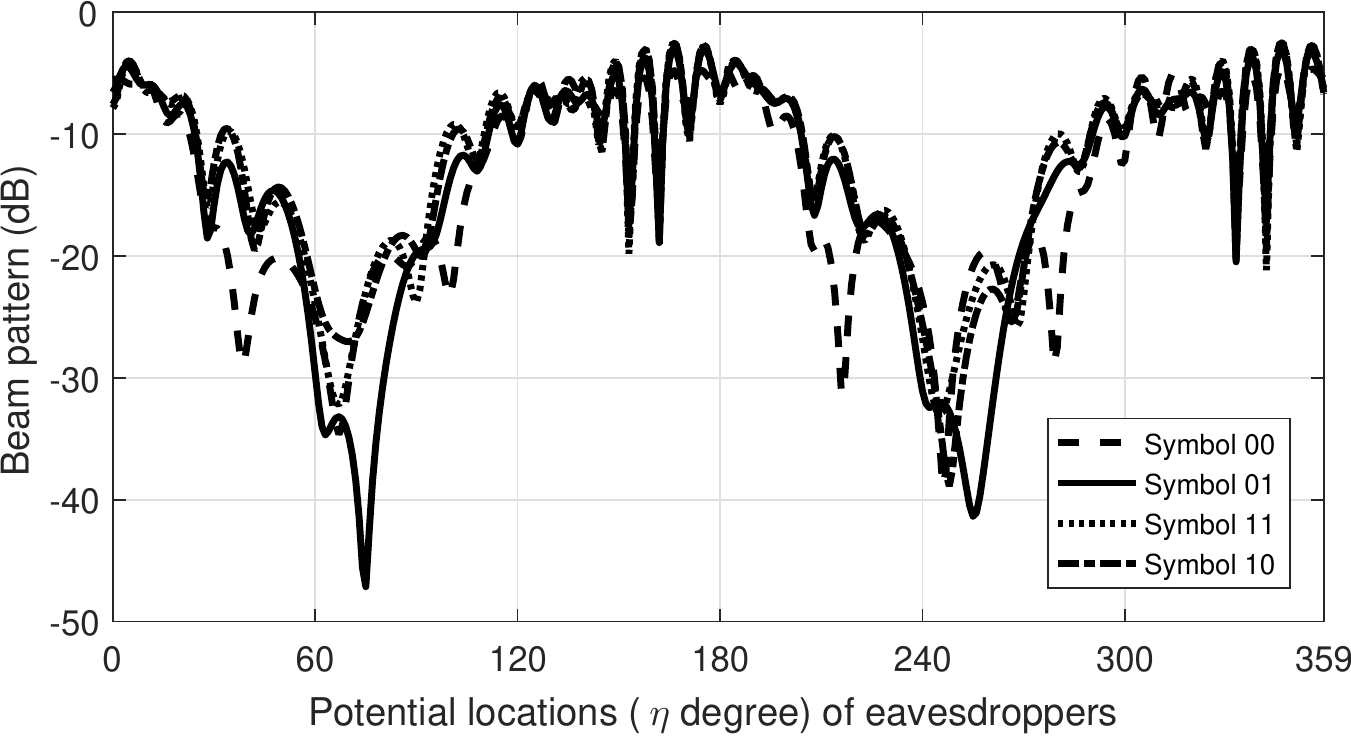}
  \label{fig:response_ula}}
  \subfigure[]{
  \includegraphics[width=0.35\textwidth]{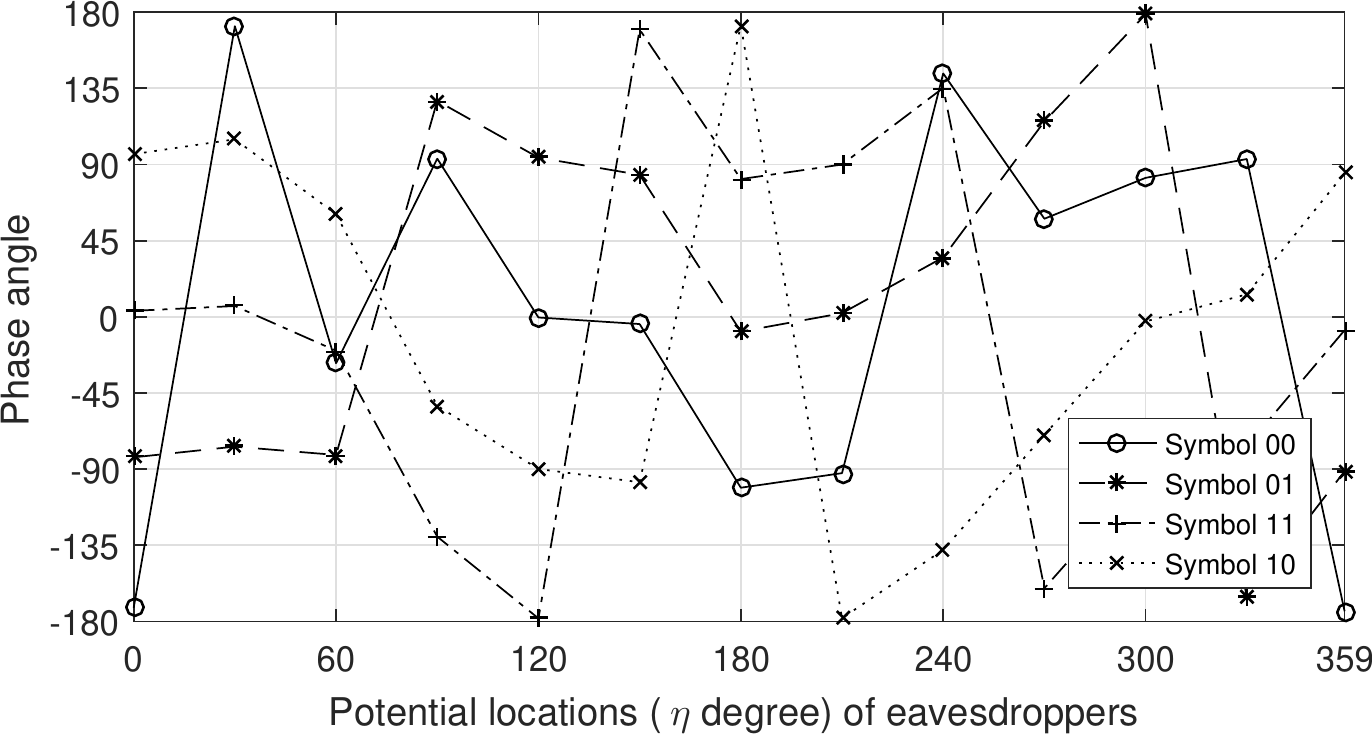}
  \label{fig:phase_ula}}
  \caption{Resultant beam and phase patterns based on the ULA design \eqref{eq:mindm} for eavesdroppers.}
\end{figure}

The resultant beam and phase patterns for the eavesdroppers based on the ULA design \eqref{eq:mindm} are shown in Figs. \ref{fig:response_ula} and \ref{fig:phase_ula}, where the beam response level at all locations of the eavesdroppers ($\eta\in[0^\circ, 360^\circ$)) is lower than $0$dB which is the beam response for the desired locations. The phase of signal at these eavesdroppers are random while the desired phase for these four symbols should be QPSK modulation, as mentioned before. The beam and phase patterns for the sparse array design in \eqref{eq:approx} are not shown as they have similar characteristics to ULA's beam and phase responses. As shown in Table \ref{tb:comparision}, with a fewer number of antennas, the sparse array design results provide a better match to the desired responses based on the error norm of array responses.

\begin{figure}
  \centering
  \subfigure[]{
  \includegraphics[width = 0.35\textwidth]{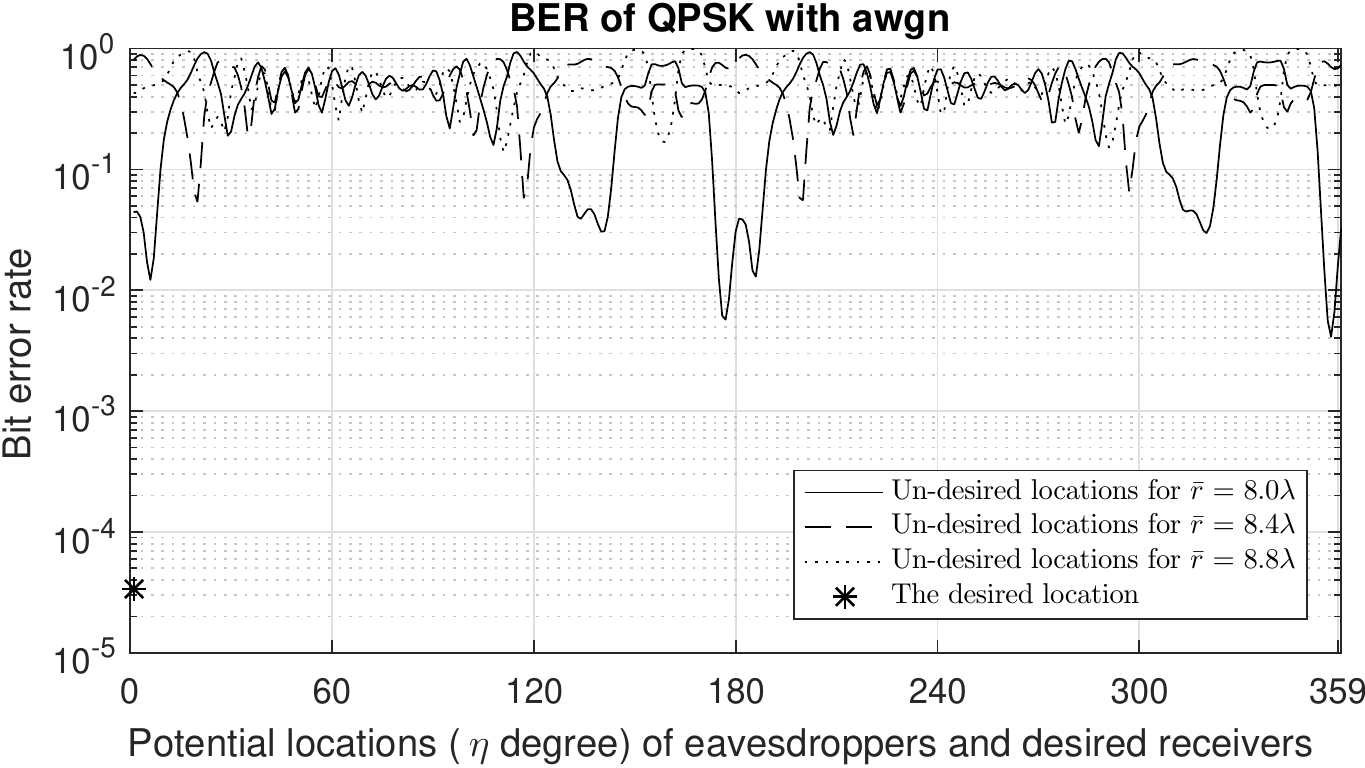}
  \label{fig:ber_ula}}
  \subfigure[]{
  \includegraphics[width=0.35\textwidth]{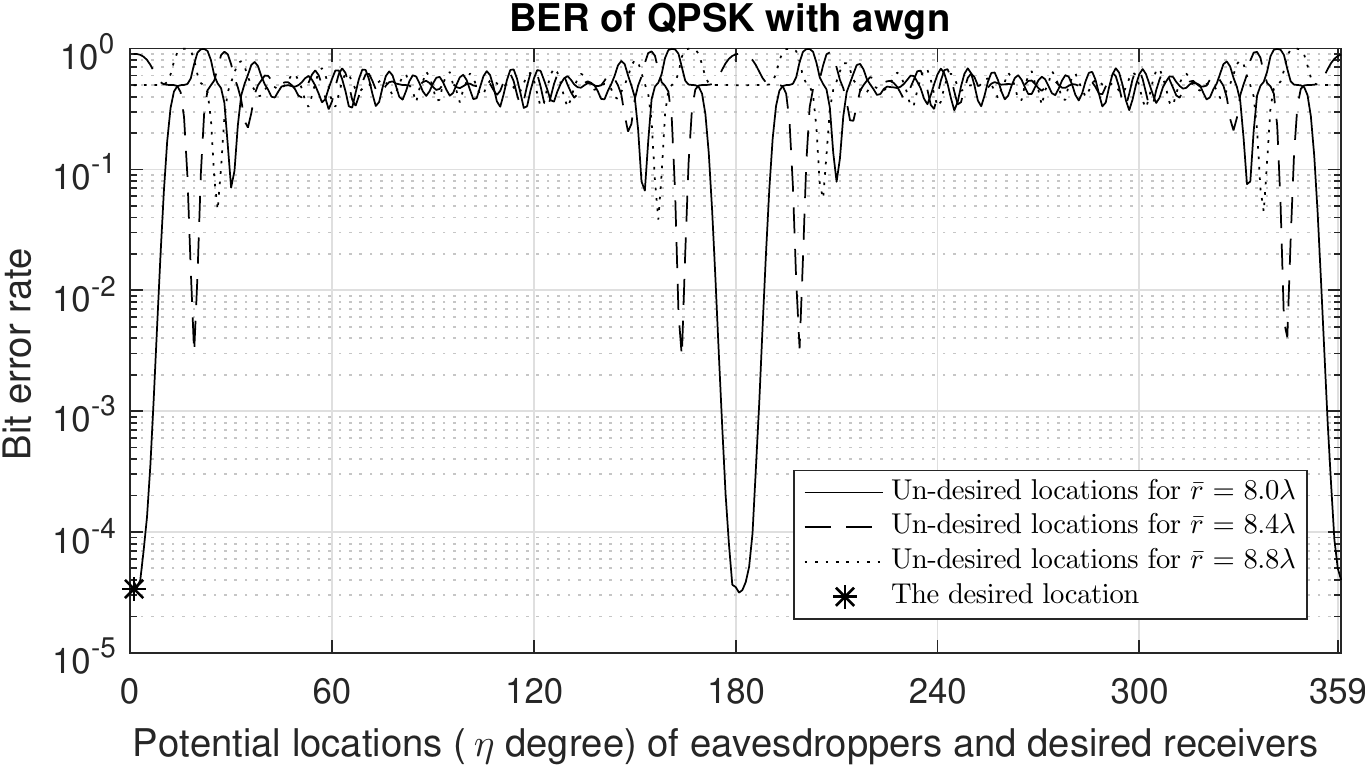}
  \label{fig:ber_ula_no_reflector_for_comparison}}
  \caption{BERs patterns for the eavesdroppers and desired receiver based on ULA designs (a) in multi-path model \eqref{eq:mindm} and (b) in LOS model.}
\end{figure}

Considering the imperfect knowledge of the geometry, e.g. the locations of eavesdroppers are not exactly the same as the locations we thought. Here we assume eavesdroppers are distributed on the circumferences of the circles with $\bar{r} = 8\lambda$ and $\bar{r} = 8.8\lambda$, while the set of weight coefficients are designed for $\bar{r} = 8.4\lambda$. Fig. \ref{fig:ber_ula} shows the BERs based on the ULA design \eqref{eq:mindm} in the multi-path model, where BERs at these eavesdroppers in these cases are still much higher than the rate in the desired location ($10^{-5}$). While in LOS model, as shown in Fig. \ref{fig:ber_ula_no_reflector_for_comparison}, BERs based on $\bar{r} = 8.4\lambda$ at some positions of the eavesdroppers are close to $10^{-3}$, lower than the counterpart ($10^{-1}$) in the multi-path model, indicated by dash line in Fig. \ref{fig:ber_ula}, demonstrating the effectiveness of the multi-path scheme. Moreover, for eavesdroppers close to the desired direction and also integer wavelengths away from the desired location, e.g. $\bar{r} = 8\lambda$, $\eta = 0^\circ$ and $\eta = 180^\circ$, the BERs reach $10^{-5}$, same as in desired locations, much lower than the BERs at these positions in the multi-path model, further demonstrating the effectiveness of the proposed positional modulation designs. The BERs for the sparse array design \eqref{eq:approx} are not shown as they have similar features to the ULA designs.

\begin{table}
\caption{Summary of the design results.} \centering
\begin{tabular*}{.45\textwidth}{@{\extracolsep{\fill}} ccccc}\hline
  &$\text{ULA}$&Usual $l_1$&Reweighted\\
\hline
Antenna number  & $30$  &$117$ &$8$\\
Aperture/$\lambda$  &$14.5$ &$20$ &$19.8$  \\
Average spacing/$\lambda$          & $0.5$  &$0.1724$ &$2.8286$\\
$||\textbf{p}_{m,E}-(\bm{\nu}_{E}\cdot e^{j\bm{\psi}_{E}}\cdot(\textbf{w}_{m}^{H}\textbf{S}_{E})$&&&\\
$+\bm{\xi}_{E}\cdot e^{j\bm{\phi}_{E}}\cdot(\textbf{w}_m^{H}\hat{\textbf{S}}_{E}))||_{2}$  &&&\\ (Error norm of array responses)&$14.0707$ &$13.1773$ &$13.9028$\\
\hline
\end{tabular*}
\label{tb:comparision}
\end{table}

\section{Conclusions}\label{sec:con}

In this paper, a two-ray transmission model has been studied for positional modulation, where signals via LOS and reflected paths are combined at the receiver side. With the positional modulation technique, signals with a given modulation pattern can only be received at desired
locations, but scrambled for positions around them. By the proposed designs, the multi-path effect is exploited to overcome the drawback of traditional DM design when eavesdroppers are aligned with or very close to the desired users. Examples for a given array geometry and an optimised sparse array have been provided to verify the effectiveness of the proposed designs.
\renewcommand\refname{\center \normalsize REFERENCES}
\bibliographystyle{my_IEEEtran}
\bibliography{mybib}
\end{document}